\documentclass[aos,preprint]{imsart}

\RequirePackage[OT1]{fontenc}

\RequirePackage{graphicx,amsthm,amsmath,natbib,lineno} 
\RequirePackage[colorlinks,citecolor=blue,urlcolor=blue]{hyperref} 
\RequirePackage{hypernat} 
\usepackage{bbm}
\arxiv{stat.ME/0908.2794}
\startlocaldefs 
\newtheorem{definition}{Definition}[section] \numberwithin{equation}{section} \theoremstyle{plain} 
\newtheorem{thm}{Theorem}[section] 
\newtheorem{proposition}{Proposition}[section] 
 
\newtheorem{example}{Example}[section] 
\newtheorem{Remark}{Remark}[section] 
\endlocaldefs 
\newcommand{\R}{\mathbbm{R}}

\begin{document} 

\begin{frontmatter}
  \title{A nonparametric independence test using random permutations
    \protect\thanksref{T1}} \runtitle{A nonparametric independence
    test } \thankstext{T1}{This work is partially supported by
    PRONEX/FAPESP Project Stochastic behavior, critical phenomena and
    rhythmic pattern identification in natural languages (grant number
    03/09930-9) and by CNPq Edital Universal (2007), project:
    ``Padr\~{o}es r\'{ \i}tmicos, dom\'{ \i}nios pros\'{o}dicos e
    modelagem probabil\'{ \i}stica em corpora do portugu\^{e}s''.}
\begin{aug}
	\author{\fnms{Jes\'{u}s} \snm{ E. Garc\'{ \i}a}\thanksref{t1,m1}\ead[label=e1]{jg@ime.unicamp.br}} \and 
	\author{\fnms{Ver\'{o}nica} \snm{ A. Gonz\'{a}lez-L\'{o}pez}\thanksref{t1,m1}\ead[label=e2]{veronica@ime.unicamp.br}}
	
	\thankstext{t1}{Departamento de Estat\'{ \i}stica. Intituto de
          Matem\'{a}tica Estat\'{ \i}stica e Computa\c c\~{a}o
          Cient\'{ \i}fica.}
	
	\runauthor{J. E. Garc\'{ \i}a and V. A. Gonz\'{a}lez-L\'{o}pez}
	
	\affiliation{Universidade Estadual de Campinas \thanksmark{m1}}
	
	\address{Departamento de Esat\'{ \i}stica\\
	Instituto de Matem\'{a}tica Estat\'{ \i}stica e Computa\c c\~{a}o Cient\'{ \i}fica\\
	Universidade Estadual de Campinas \\
	Rua Sergio Buarque de Holanda,651\\
	Cidade Universit\'{a}ria-Bar\~{a}o Geraldo\\
	Caixa Postal: 6065\\
	13083-859 Campinas, SP, Brazil\\
	\printead{e1}\\
	\phantom{E-mail:\ }\printead*{e2}}
	
	%
\end{aug}
\begin{abstract}
  We propose a new nonparametric test for the supposition of
  independence between two continuous random variables $X$ and $Y.$ Given a sample of $(X,Y),$ the test is
  based on the size of the longest increasing subsequence of the
  permutation  which maps the ranks of the $X$ observations to the ranks of the $Y$ observations. We identify the independence assumption between the
  two continuous variables with the space of permutation equipped with
  the uniform distribution and we show the exact distribution of the
  statistic. We calculate the distribution for several sample
  sizes. Through a simulation study we estimate the power of our test
  for diverse alternative hypothesis under the null hypothesis of
  independence.
\end{abstract}
\end{frontmatter}

\section{Introduction} Let $(X,Y)$ be a random vector of continuous
variables with unknown joint cumulative distribution $H$ and univariate marginal distributions $F$ and
$G.$ Call $\Omega$ the space of the univariate, cumulative and continuous distributions, then $F, G \in \Omega$.

 Suppose that
$(x_1,y_1),\cdots ,(x_n,y_n)$ is a paired sample of size $n$ of
$(X,Y).$ We want to test the hypothesis 
\begin{eqnarray}\label{testehyp1}
H_0: X\,\mbox{and}\,Y\,\mbox{are independent}. 
\end{eqnarray}
A test is constructed with no extra assumption (other than continuity) about the form of the
marginal distributions.  Let $rank(x_i) \; (rank(y_i))$ be the position occupied by $x_i \; (y_i)$ in the sample $\left\{ x_j \right\}_{j=1}^n (\left\{ y_j \right\}_{j=1}^n),$ the test statistic depends on the rank order of the observations.
The procedure is based on the size of the longest increasing subsequence of the random permutation defined by the paired samples. \\
The power of our test is compared with those of various existing tests by simulation. Four independence tests are selected for this comparative process, namely, Pearson test, Kendall test, Spearman test and Hoeffding test. One of them is parametric, the Pearson test, selected by its well known performance in the normal case and the other three are nonparametric. Each methodology estimates the association between paired samples and computes a test of the value being zero. They use different measures of association, all of them in the interval $[-1,1]$ with 0 indicating no association (depending on the test's formulation). \\
In our simulations, the Hoeffding test has a better power but at the expense of not controlling the significance level.  In general lines, in the independent non normal marginals case, for moderate sample size, our test is the only one respecting the significance level. 
On the other hand, in the dependent case, the performance of our test depends on the joint distribution. Assuming normal joint distribution and linear dependence between the normal random variables, our test performs a lower power compared to the other tests which are designed for that case. For the case in which the joint distribution is not normal, we performed a simulation study with different conditions. For example, we use a mixture of bivariate normal distributions on the random variables. In that case our procedure was competitive and more powerful than the other four tests considered. In these simulations just our procedure and Hoeffding had a power function going to 1 when the sample size grows.  \\
Section \ref{principal} is devoted to motivate the proposal and
provides the main concepts and the definition of the test
statistic. In Section \ref{themain} we show how to calculate the exact
distribution and the asymptotic distribution of the test statistic. In
Section \ref{aplicar} we show the effectivity of our proposal, using
simulations and we discuss the results. The
Appendix \ref{backg} contains a brief overview of the tests that we
use to compare with our proposal.
 
\section{Nondecreasing (nonincreasing) subsets}\label{principal} In order to highlight the relationship between the values observed of $X$ and $Y,$ we can plot $X$ versus $Y,$ detecting in some specific cases evidence about the form of the function $g$ connecting the variables. In this way the functional relation $Y=g(X)$ could be established. The specification of the form of $g$ is in general a hard task. Instead of looking for $g$ directly, we can ask for which kind of random bivariate distribution $H$ assures that $Y$ is almost surely an increasing (or decreasing) function of $X.$ The answer is independent of the marginal distributions $F$ and $G,$ if $F$ and $G$ are in $\Omega .$ 
\subsection{Perfect dependence} 
\begin{definition}
If $\overline{\R}=[-\infty,+\infty] $ and $\overline{\R}^2=\overline{\R}\times \overline{\R},$ 
\begin{itemize}
	\item[1.]a subset $S$ of $\overline{\R}^2$ is nondecreasing if for any $(x,y)$ and $(u,v)$ in $S,$ $x<u$ implies $y\leq v$ (see figure \ref{fig:nondecreasingsubset}); 
	\begin{figure}
		[h] 
		
		\centering 
		\includegraphics[width=2.5in]{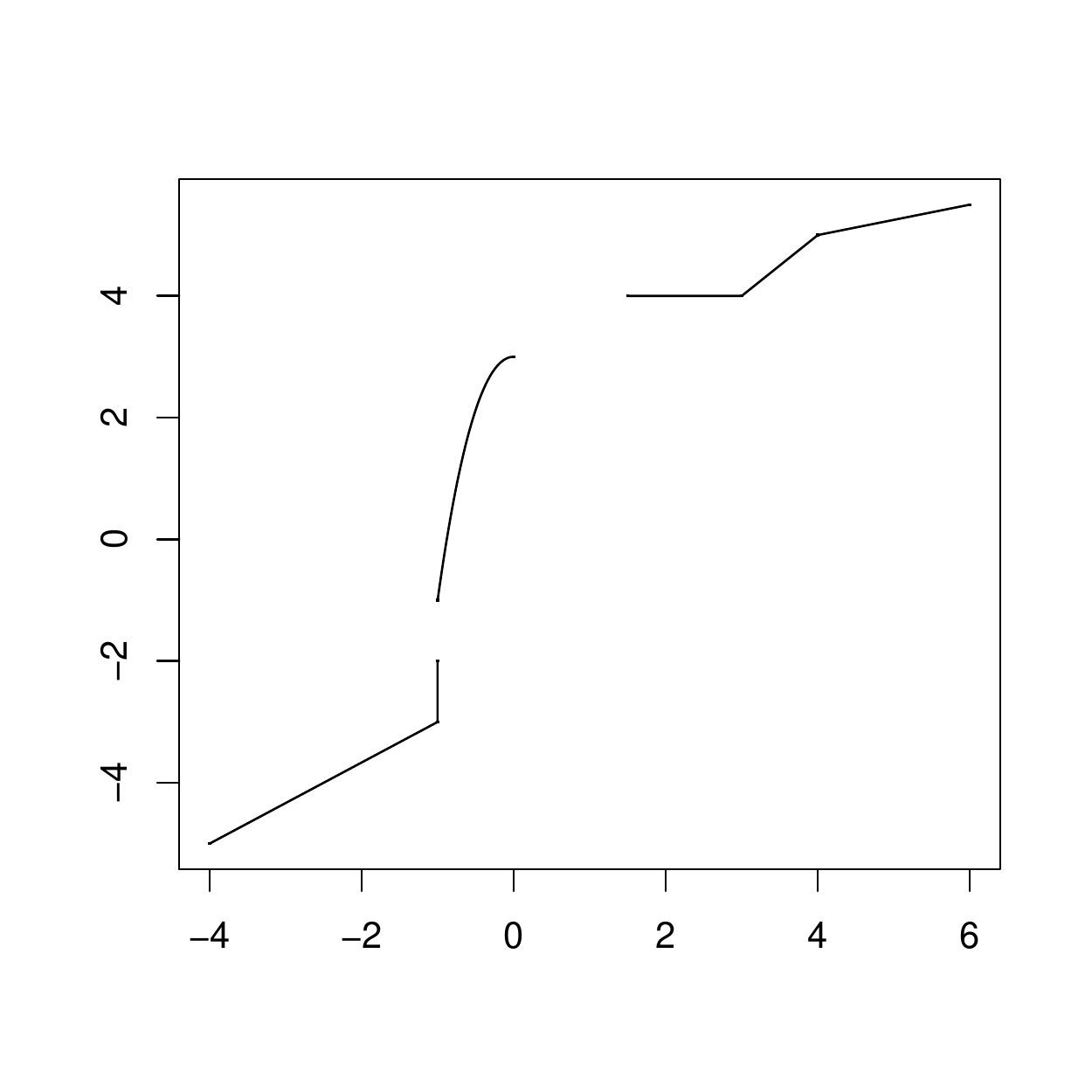} \caption{The graph of a nondecreasing set.} \label{fig:nondecreasingsubset} 
	\end{figure}
	\item[2.]a subset $S$ of $\overline{\R}^2$ is nonincreasing if for any $(x,y)$ and $(u,v)$ in $S,$ $x<u$ implies $y \geq v.$ 
\end{itemize}
\end{definition}
We refer to 
\begin{eqnarray}
H(x,y)=\min \left\{F(x),G(y)\right\} \label{min}\\
H(x,y)=\max \left\{0,F(x)+G(y)-1\right\} \label{max}
\end{eqnarray}
 as the Fr\'{e}chet upper bound and the Fr\'{e}chet lower bound respectively.

The next theorem establishes that $H$ is identically equal to its Fr\'{e}chet upper (lower) bounds if and only if the support of $H$ is concentrated on a nondecreasing (nonincreasing) subset. 
\begin{thm}
\citet{Mikusinski1991}.\label{incresubset} Let be $H$ the joint distribution of a pair $X, Y$ of random variables whose one dimensional distribution functions are $F$ and $G,$ respectively. Then, 
\begin{itemize}
	\item[1.] $H(x,y)$ is identically equal to (\ref{min}) if and only if $(X,Y)$ lies almost surely in a nondecreasing subset of $\R^2;$ 
	\item[2.]$H(x,y)$ is identically equal to (\ref{max}) if and only if $(X,Y)$ lies almost surely in a nonincreasing subset of $\R^2.$ 
\end{itemize}
\end{thm}
If $X$ and $Y$ are continuous, the support of $H$ can have no vertical or horizontal lines. When $H(x,y)$ is given by (\ref{min}) (or (\ref{max})) $X$ and $Y$ are continuous, $Y$ is almost surely an increasing (decreasing) function of $X.$ \\
Every kind of dependence is found between two pure cases of dependence, monotone nonincreasing and monotone nondecreasing as showed by the next proposition. 
\begin{proposition}
\citet{Nelsen1999}. Consider the same hypotheses as in Theorem \ref{incresubset} . Then, 
\begin{itemize}
	\item[1.]$ \max\left\{0,F(x)+G(y)-1 \right\} \leq H(x,y) \leq \min\left\{F(x), G(y) \right\}\,\,\,\forall x,y \in \R ;$ 
	\item[2.]$ \max\left\{0,u+v-1 \right\} \leq C(u,v) \leq \min\left\{u, v \right\},\,\,u,v\in [0,1],$ where $C$ is a cumulative distribution (or copula) such that $H(x,y)=C(F(x),G(y)).$ 
\end{itemize}
\end{proposition}
\begin{Remark}
As showed in the last result, the dependence between $X$ and $Y$ is exposed transforming the variables $X$ and $Y$ by the marginal cumulative $F$ and $G,$ respectively. Under the continuous marginal suppositions, if $H(x,y)$ is given by (\ref{min}) (or (\ref{max})), then $P(U=V)=1 \,\, (\mbox{or}\,\,P(U=1-V)=1),$ where $U=F(X)$ and $V=G(Y).$ 
\end{Remark}
In conclusion, one way to expose the dependence, with little information about the marginal behavior of $X$ and $Y,$ is to use the empirical marginal distribution where each marginal observation $x_i (y_i)$ is replaced by $\frac{rank(x_i)}{n} (\frac{rank(y_i)}{n}).$\\
Our proposal consists on showing a specific relationship between $X$ and $Y$ which makes easy to measure the independence between them. We show the relation induced by the empirical copula, replacing the original observations by its marginal ranks and we find the longest increasing subsequence defined by the graphic of the marginal ranks. First, we note that the distribution of the statistic given by the longest increasing subsequence is known under the assumption of independence. Second, the longest increasing subsequence exposes the tendency of the data to accumulate points into the increasing subset defined by the longest increasing subsequence. 

\subsection{Construction of a nondecreasing subset using the sample} We connect the sample with a specific permutation of $n$ points $\pi_s,$ this permutation defines the nondecreasing subset that we use. We explain the procedure using the next warm-up example. 
\begin{example}
\label{ex2} Let us consider the random sample $s,$ $$\left\{(4.16,3.25),(1.15,3.5),(2.51,4.17),(3.61,3.18),(1.81,2.86)\right\}.$$ First, sort the samples $\left\{(x_i,y_i)\right\}_{i=1}^n$ in increasing order in relation to the sample $\left\{ x_i \right\}_{i=1}^n$ and replace the $x_i$ value with its rank in the sequence, on our example this produces $\left\{(1,3.5),(2,2.86),(3,4.17),(4,3.18),(5,3.25)\right\}.$ Next, replace each $y_i$ with its rank in the $\left\{ y_i \right\}_{i=1}^n$ sequence, on our example this produces $\left\{(1,4),(2,1),(3,5),(4,2),(5,3)\right\}.$ The permutation $\pi_s$ related to this sample is defined by $$\pi_s(1)=4, \pi_s(2)=1, \pi_s(3)=5, \pi_s(4)=2, \pi_s(5)=3.$$ 
\begin{figure}
	[h] 
	
	\centering 
	\includegraphics[width=5.4in]{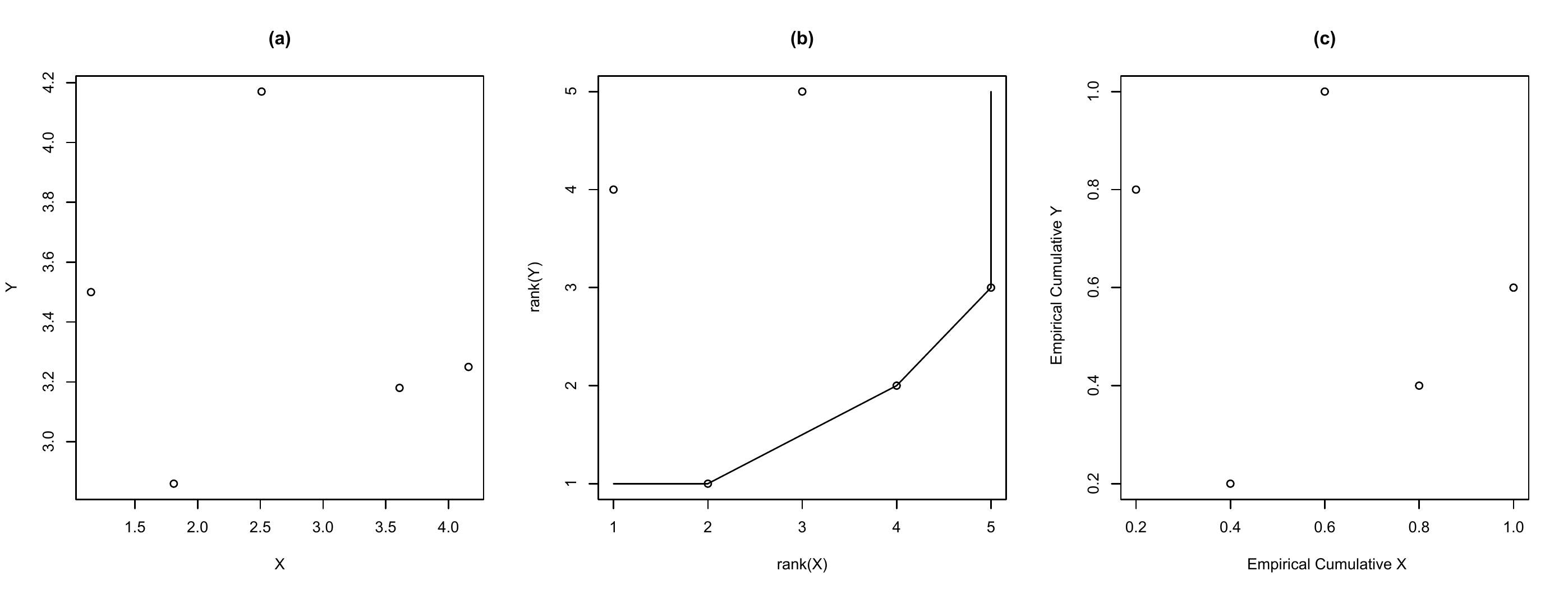} \caption{Dispersion's graphic and permutation (Example \ref{ex2}). (a) is the dispersion plot for the sample. (b) represents the permutation defined by the sample, the black line shows the longest increasing subsequence. (c) shows the empirical copula of the sample. } \label{fig:example} 
\end{figure}

On this example the longest increasing subsequence is $\left\{1,2,3\right\}$ see figure \ref{fig:example} (b). 
\end{example}

Our test is based on the distribution of the size of the longest increasing subsequence of a random permutation of $n$ points, assuming uniform distribution on the random permutation space. \\

Formally, 
\begin{definition}\label{increasingpi} 
Let $\mathcal{S}_n$ denote the group of permutations of $\left\{1,\cdots,n \right\}.$ If $\pi \in \mathcal{S}_n,$ we say that $\pi(i_1),\cdots, \pi(i_k)$ is an increasing subsequence in $\pi$ if $1\leq i_1 < \cdots < i_k \leq n$ and $1 \leq \pi(i_1)<\pi(i_2)<\cdots <\pi(i_k) \leq n.$
\end{definition}

\begin{definition}
\label{defLn} Given a random permutation $\pi \in \mathcal{S}_n,$ 
 \begin{itemize}
\item[1.] we call $L_n(\pi)$  the length of the longest increasing subsequence of $\pi ;$\\
\item[2.] we  call $LD_n(\pi)$ the length of the longest decreasing subsequence of $\pi .$
\end{itemize} 
\end{definition}
\begin{example}
\label{ex1} Consider the set $\left\{ 1,2,3,4,5,6,7,8\right\}.$ Let  $\pi$ be the permutation which transforms the previous set in  $\left\{ 3,6,1,7,4,2,5,8\right\}$ where $\pi(1)=3,\pi(2)=6,\pi(3)=1,\pi(4)=5,\pi(5)=7,\pi(6)=2,\pi(7)=4,\pi(8)=8.$ Examples of increasing subsequences are $\left\{ 1,7,8 \right\}$, $\left\{ 3,6,7,8 \right\}$, $\left\{ 1,2,5,8 \right\}.$ The maximal size for the increasing subsequences is $4$ which is reached by the sequences $\left\{ 1,2,5,8 \right\}$, $\left\{ 1,4,5,8 \right\}$ and $\left\{ 3,6,7,8 \right\}$, then $L_8(\pi)=4.$ 
\end{example}
\begin{example}
(continued). On the Example \ref{ex2} the longest increasing subsequence is $\left\{1,2,3\right\}$ and the value $L_5(\pi_s)=3,$ see figure \ref{fig:example}. 
\end{example}
On the next section we study the distribution of the length of the longest increasing subsequence, under the assumption of independence between $X$ and $Y.$ 

\section{The longest increasing subsequence}\label{themain} Let $\mathcal{S}_n$ denote the group of permutations of $\left\{1,\cdots,n \right\}$ and equip $\mathcal{S}_n$ with the uniform distribution, for $k=1,2\cdots,n,$ we define, 
\begin{equation}
\label{Lnallpi} P(L_n=k)=\frac{\#\left\{ \,\, \mbox{of permutations}\,\,\, \pi \in \mathcal{S}_n: L_{n}(\pi)=k \right\} }{n!}.
\end{equation}
We denote \ref{Lnallpi} briefly by $p_k^n.$\\
Under the independence hypothesis for the random variables $X$ and $Y,$ every possible permutation defined by a random sample of size $n,$ $\left\{(x_i,y_i)\right\}_{i=1}^n$, has the same probability $1/n!$. Using this fact, the Young tableaux, the Schensted theorem by \citet{Schensted1961}  and the  {\it ZS2} algorithm by \citet{ZS2}, the probabilities $p_k^n$ could be calculated for each finite $n$ and and $k$ with $1 \leq k\leq n$. 

\subsection{The exact distribution of $L_n$ in the case of independence} We will touch only a few aspects of the theory, just the necessary in order to show how to calculate the distribution.
Firstly, we introduce the same concepts. 
\begin{definition}
A standard Young Tableau of order $n$ is an arrangement of $n$ distinct natural numbers in rows and columns so that the numbers in each row and in each column form increasing sequences, and so that there is an element of each row in the first column and an element of each column in the first row, and there are no gaps between numbers. 
\end{definition}

The first row on the standard Young Tableau corresponds to one of the longest increasing subsequences. We can construct the Young tableaux composed by the increasing subsequences originated by some specific permutation, as showed in the next example. 
\begin{example}
\label{ex3}(continued). For the permutation on Example \ref{ex2} the Young tableau is, 
\begin{eqnarray*}
	1&2&3\\
	4&5 
\end{eqnarray*}
\end{example}
\begin{definition}
If $T$ is a standard Young tableau of order $n,$ for each element $j,\,\,j \in \left\{ 1,\cdots ,n\right\}$ of the arrangement we define the Hook number of $j$ as the number of elements in the same column and in the same row in which $j$ is included. Counting from the bottom until the element $j$ and from the right to the row until the element $j.$ 
\end{definition}
\begin{example}
(continued). For the standard Young tableau in Example \ref{ex3} the Hooks numbers are, 
\begin{eqnarray*}
	4&3&1\\
	2&1 
\end{eqnarray*}
\label{exhook} 
\end{example}
\begin{Remark}
The Hook numbers depend on the form of the Tableau not on the numbers filling it. Different permutations of $\left\{1, \cdots , n\right\}$ can give the same Tableau shape, so, each permutation is directly associated with the shape of a Young tableau.
\end{Remark}
The next example shows all the possible shapes of the Young tableaux that can be obtained by the permutations of 5 numbers.
\begin{example}\label{listshapes}
The complete list of shapes and Hooks numbers in each shape, admitted by the numbers $\left\{ 1,2,3,4,5\right\}$ follows in the next table. Each element of this list is associated with an integer partition (IP) of $n=5.$\\

{\footnotesize{
\begin{tabular}{l||l||l||l||l||l||l}
Shape 1&Shape 2 & Shape 3 & Shape 4 & Shape 5 & Shape 6 & Shape 7\\ \hline
5 & 51&52&521&431&5321&54321\\
	4&3&31&2&21&1&\\
	3&2&1&1&&&\\
	2&1&&&&&\\
	1 &&&&&&\\ \hline \hline
	IP1(5)& IP2(5)&IP3(5)&IP4(5)&IP5(5)&IP6(5)&IP7(5)\\
	5 & 4+1& 3+2&3+1+1&2+2+1&2+1+1+1&1+1+1+1+1
\end{tabular}
}}\\

Shape 1 corresponds to the permutation $\pi(1)=5, \pi(2)=4, \pi(3)=3, \pi(4)=2, \pi(5)=1$ ($LD_5=5$) and it is associated to the integer partition of $n=5,$ IP1(5)=5 (the sum of the number of elements in the first column of the shape 1). The shape 5 is associated to the integer partition of $n=5,$ IP5(5)=2+2+1, where each term (from left to right) is the sum of the number of elements by column in the shape 5.
\end{example}
In order to calculate all the possible shapes of Young tableaux of size $n,$ having $k$ columns and $m$ rows we use an algorithm which finds these forms (or the integer partitions) in an efficient way.

Given a permutation $\pi,$ the size of the longest increasing subsequences for $\pi$ is the size of the first row in the shape of the Tableaux corresponding to the permutation. In other words, the number of permutations $\pi$ of $n$ numbers such that $L(\pi)=k$, is the number of Young tableaux with a shape such that the first row has size $k$. The number of standard Young tableaux with a given shape can be efficiently computed using the following theorem by \citet{Frame1954}. 
\begin{thm}
\citet{Frame1954} \label{NYT}. The number of standard Young tableaux with a given shape, containing the integers $\left\{1, \cdots , n\right\}$ is $\frac{n!}{\prod_{j=1}^{n}h_j}$ where the $h_j,j=1,\cdots,n$ are the Hook numbers associated with the cells of the Tableau. 
\end{thm}
\begin{example}
(continued). The number of standard Young tableaux containing the numbers $\left\{1,2,3,4,5\right\}$ with shape given by the Example \ref{exhook} is $5!/[4.3.2]=5.$ 
\end{example}
The number of sequences of size $n$ with a longest increasing subsequence of size $k$ and longest decreasing subsequence of length $m$ can be calculated using the result given by \citet{Schensted1961}. 
\begin{thm}
\citet{Schensted1961}\label{PLn}. The number of sequences consisting of the numbers $\left\{1,\cdots ,n\right\}$ and having a longest increasing subsequence of length $k$ and longest decreasing subsequence of length $m$, is the sum of the squares of the number of standard Young tableaux of identical shape, having $k$ columns and $m$ rows. 
\end{thm}
\begin{example} \label{ex232} (continued from example \ref{listshapes}).
Considering the numbers $\left\{ 1,2,3,4,5\right\}$ we want to calculate the number of sequences having $L_5=3.$ Let us denote by $\#\left\{ A \right\}$ the cardinal of $A,$ where $A$ is some set.\\
$\#\left\{L_5=3\right\}=\#\left\{L_5=3,LD_5=2\right\}+\#\left\{L_5=3,LD_5=3\right\},$ corresponding with only two possible shapes of Young tableaux, with Hook numbers given by the example \ref{listshapes}, shape 4 and shape 5.
Using the Schensted Theorem,\\
$\#\left\{L_5=3,LD_5=2\right\}=5^2=25,\,\,\#\left\{L_5=3,LD_5=3\right\}=6^2=36$ and $\#\left\{L_5=3\right\}=25+36=61.$ 
\end{example}

For a given shape $W,$ we call $N(W)$ the number of standard Young tableaux with that shape as given by Theorem \ref{NYT}. Let $V_n(k,m)$ be the set of shapes of Young tableaux of size $n$ having $k$ columns and $m$ rows. From Theorem \ref{PLn}, we have that the number of permutations of $n$ elements with a longest increasing subsequence of size $k$ and longest decreasing subsequence of length $m$ is, $$\sum_{W \in V_n(k,m) } N(W)^2$$ and the number of permutations of $n$ elements with a longest increasing subsequence of size $k$ is $$\sum_{k=1}^{n}\;\; \sum_{W \in V_n(k,m) } N(W)^2 $$ so we have the following theorem. 
\begin{thm}
\label{PKN} Let $\mathcal{S}_n$ denote the group of permutations of $\left\{1,\cdots,n \right\}$ with the uniform distribution. Let $L_n (\pi)$ be given by definition \ref{defLn} and $p_k^n,\,\,k=1,\cdots ,n$ given by the equation \ref{Lnallpi}. Then, 
\begin{equation} \label{Lnprob2}
p_k^n=\frac{1}{n!}\sum_{k=1}^{n} \sum_{W \in V_n(k,m) } N(W)^2.
\end{equation}
\end{thm}
 
There are diverse algorithms in the literature to find $V_n(k,m)$, we implemented the {\it ZS2} algorithm by \citet{ZS2}.

Using Theorem \ref{PKN} we compute $p_k^n$ for $1\leq k \leq n$, $n=1, \cdots,100.$ The table can be accessed from our R package LIStest.
\subsection{The asymptotic distribution of $L_n$ in the case of independence}The asymptotic distribution for random permutations, after appropriate centering and scaling, was first obtained by  \citet{Baik1999}, as shows the next theorem. Let $q(z)$ denote the solution of the Painlev\'{e} II equation given by,
\begin{eqnarray*}
q_{zz}= 2q^3+zq, \,\,\mbox{satisfying the boundary condition}
\end{eqnarray*}
\begin{eqnarray*}
q(z) \sim Ai(z)\,\,\,\mbox{when}\,\, z\to \infty
\end{eqnarray*}
where $Ai$ is the Airy function.
\citet{Hast1980} show the asymptotic solutions,
\begin{eqnarray*}
q(z)= -Ai(z) + O(\frac{e^{-(4/3)z^{3/2}}}{z^{1/4}}) \,\,\mbox{as}\,\,\,\, z \to \infty,\\
q(z)= - \sqrt{\frac{-z}{2}}(1+ O(\frac{1}{z^2}))\,\,\mbox{as}\,\,\,\, z \to -\infty.
\end{eqnarray*} 

Now, we  define the Tracy-Widom distribution by the next cumulative distribution,
\begin{eqnarray}\label{TWbeta2}
F_{TW}(t)=\exp\Big(-\int_{t}^{\infty}(z-t)q^2(z)dz\Big), \,\,t \in \R .
\end{eqnarray}
\begin{thm} \citet{Baik1999}. Let $\mathcal{S}_n$ denote the group of permutations of $\left\{1,\cdots,n \right\}$ with the uniform distribution. Let $L_n (\pi)$ be given by definition \ref{defLn}. 
Let $\chi$ be a random variable whose distribution function is $F_{TW},$ given by equation \ref{TWbeta2}.  Then, as $n \to \infty,$
\begin{eqnarray*}
\chi_n= \frac{L_n-2\sqrt{n}}{n^{1/6}} \to \chi\,\,\,\mbox{in distribution}.
\end{eqnarray*}
\end{thm}
We calculate the quantiles of the Tracy Widom distribution, using the S-plus code available in  http://www.vitrum.md/andrew/MScWrwck/codes.txt. See table \ref{QTW} for a few values.

\begin{table}[htdp]
\caption{Quantiles for the Tracy-Widom distribution}
\begin{center}
\begin{tabular}{c|c|c}
$\alpha$  & $\alpha/2$ quantile & $(1-\alpha/2)$ quantile \\ \hline
0.001 & -4.44025 & 1.54089\\
0.01& -3.91393 & 0.74618\\
0.05 & -3.44277& 0.09153
\end{tabular}
\end{center}
\label{QTW}
\end{table}%

\subsection{The $L_n$ test of independence} Let $(x_1,y_1),\cdots ,(x_n,y_n)$ be a paired sample of size $n$ of $(X,Y),$ where $X$ and $Y$ are continuous random variables with cumulative marginal $F$ and $G$ respectively; $F,\,\,G \in \Omega.$ A test for independence can be carried out, as pointed in this section.\\
The two-sided statistical tests and $P$-values are well defined when the test statistic has a symmetric distribution, which is not our case. For the asymmetric case, the most recent contributions include several proposals. We choose to use the doubled two sided P-value because it appears to be simple as a starting point.
\begin{definition}
The doubled two-sided $P$-value is given by, 
\begin{eqnarray}
\min\left\{ 2F_{L_n}(l_0)I_{ \left\{  l_0 \leq M_0 \right\}  }+2(1-F_{L_n}(l_0))I_{ \left\{ l_0>M_0 \right\} },1\right\}
\end{eqnarray}
where $l_0$ is the observed value of $L_n$ in the sample, $F_{L_n}$ is the cumulative distribution function,
$F_{L_n}(l_0)=\sum_{k=1}^{l_0}p_k^n$ (see equation \ref{Lnprob2})
and $M_0$ is the mode of the distribution. $I_{E}$ denotes the indicator function of $E.$
\end{definition} 
The previous definition was used for $n=1,\cdots, 100.$ \\  

For $n >100$ we use the asymptotic distribution of $L_n$ (see equation \ref{TWbeta2}) and the quantiles from table \ref{QTW}. 
If $\alpha \in (0,1)$ is the level of significance, we reject the hypothesis of independence, \ref{testehyp1} if $\frac{l_0-2 n^{1/2}}{n^{1/6}}< q_{\alpha/2}$ or $\frac{l_0-2 n^{1/2}}{n^{1/6}}>q_{(1-\alpha/2)},$ where $q_{\gamma}$ is the $\gamma$ quantile of $F_{TW}.$

\section{Simulation}\label{aplicar} 

To compare the power of our test against the Hoeffding, Kendall, Pearson and Spearman test (see Appendix \ref{backg}), we carried out a simulation study in which for each test we estimate the power function for different sample sizes and diverse joint distributions. For each joint distribution and sample sizes $20, 40, 60, 80, 100, 500, 1000$ we simulated $10000$ samples, and computed the $P$-values.

\subsection{Independence}
The independence case was tested in several situations.  We analize pairs of independent random variables with standard normal marginal distributions, Pareto marginal distributions, Weibull marginal distributions and Student-t marginal distributions. 

\begin{figure}
[h!] 
\centering 
\includegraphics[width=4in,height=3in]{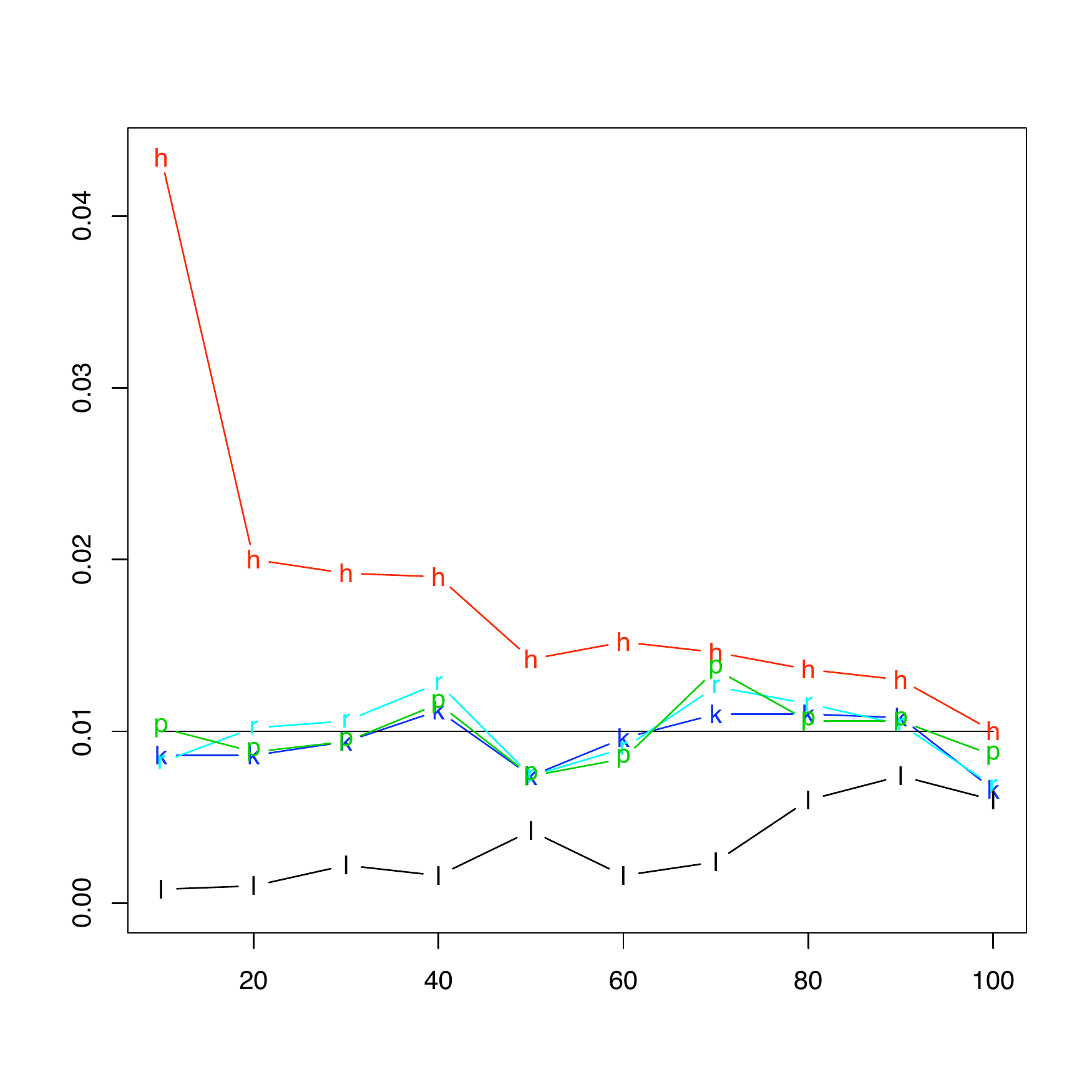} \caption{Sample size vs. empirical power function at level 0.01 in the case: independent $N(0,1)$ random variables. Hoeffding (``h'' in red); Pearson (``p'' in green); Spearman (``r'' in sky); Kendall (``k'' in blue); $L_n$ (``l'' in black).   } \label{fig:indep.norm} 
\end{figure}

Figure \ref{fig:indep.norm} shows the behavior of the empirical power functions, under independence when $X$ and $Y$ have standard normal marginals. The power function of the statistic $L_n$ (equation \ref{Lnallpi}, equation \ref{Lnprob2}) is compared with other power functions, given by Hoeffding, Pearson, Spearman and Kendall test. The power function of our test is smaller than the significance level, we can see also how the empirical power function for the Hoeffding test is not lower than the significance level. Table \ref{tab:indep.norm} shows the power for level $0.05$.

\begin{table}
[h!]
\begin{center}
\begin{tabular}{rrrrrr} 
  \hline
 & Spearman & Kendall & Hoeffding & Pearson & $L_n$  \\
  \hline
10 & 0.051 & 0.049 & 0.110 & 0.050 & 0.022  \\
  20 & 0.047 & 0.044 & 0.070 & 0.048 & 0.012  \\
  30 & 0.048 & 0.049 & 0.069 & 0.047 & 0.016  \\
  40 & 0.058 & 0.056 & 0.072 & 0.055 & 0.016  \\
  50 & 0.051 & 0.051 & 0.062 & 0.049 & 0.025  \\
  60 & 0.048 & 0.048 & 0.060 & 0.046 & 0.008  \\
  70 & 0.057 & 0.054 & 0.063 & 0.058 & 0.012  \\
  80 & 0.053 & 0.052 & 0.057 & 0.055 & 0.023  \\
  90 & 0.046 & 0.049 & 0.055 & 0.049 & 0.025  \\
  100 & 0.047 & 0.048 & 0.054 & 0.049 & 0.021  \\
   \hline
\end{tabular}
\caption{Empirical power function at level 0.05. Independent $N(0,1)$ case. }
\label{tab:indep.norm}
\end{center}
\end{table}

\begin{figure}
[h!]
\centering 
\includegraphics[width=2.3in,height=2.5in]{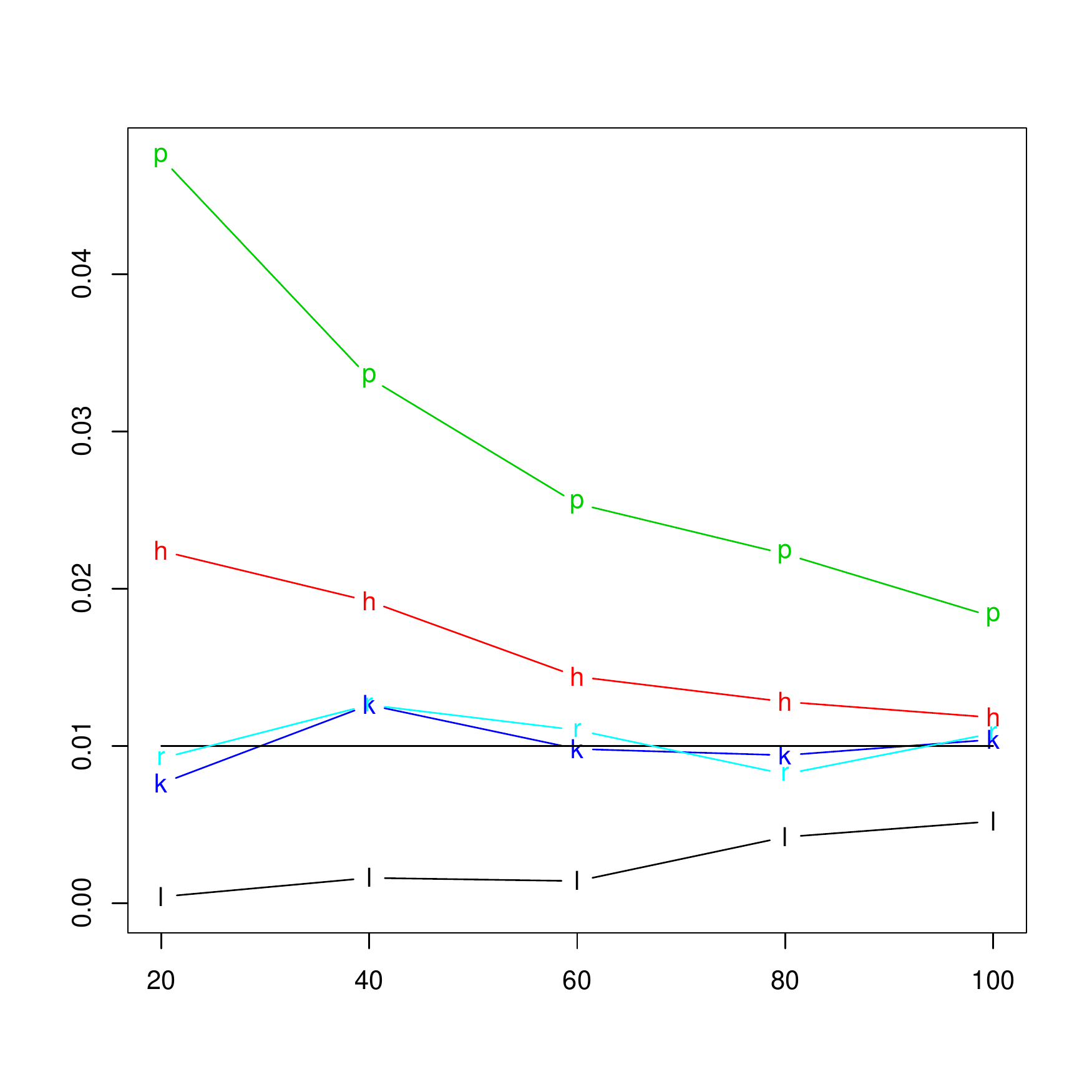}
\includegraphics[width=2.3in,height=2.5in]{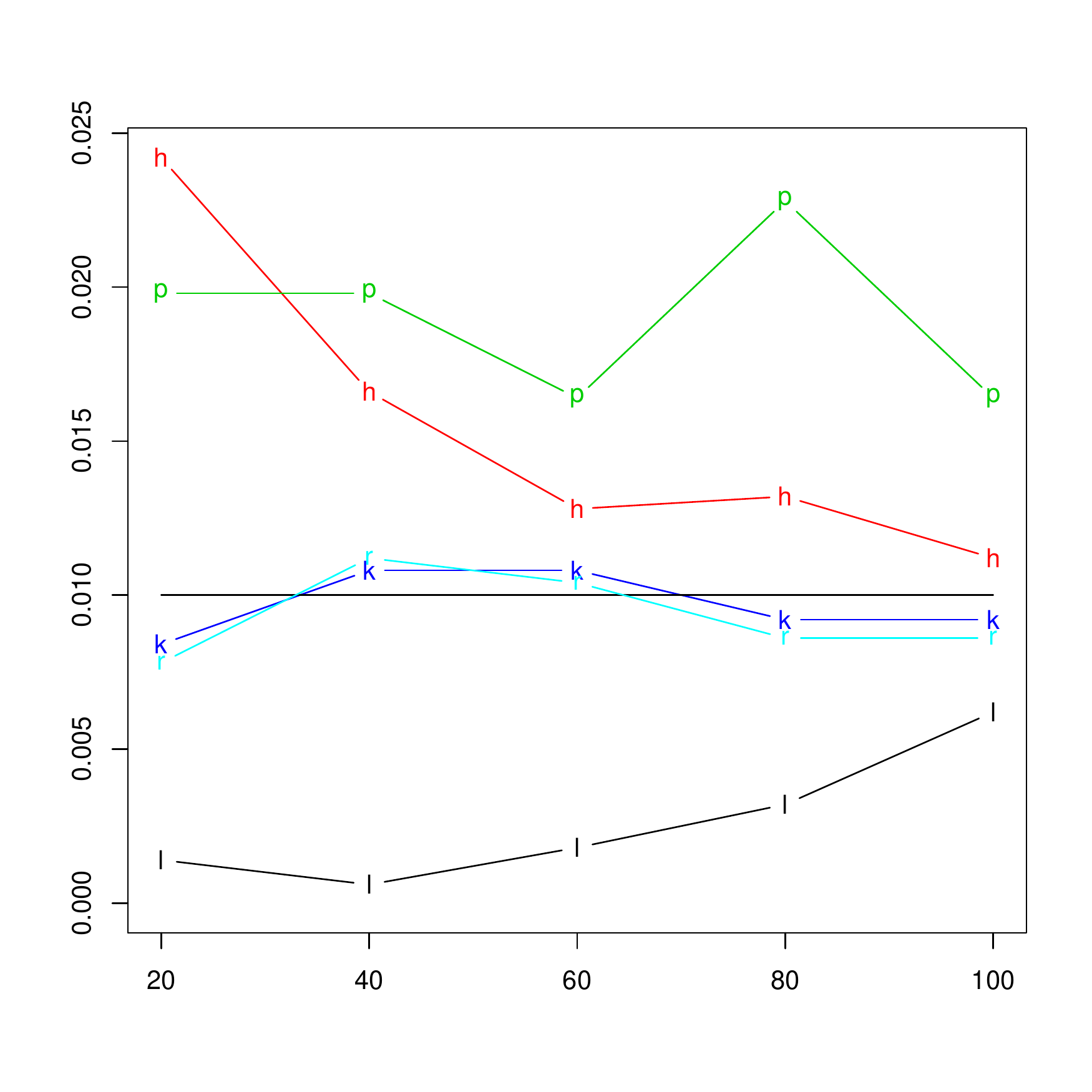}
 \caption{Sample size vs. empirical power Function at level 0.01 in the case of independent Pareto random variables with parameters $(1,0.25)$ and $(1,4)$ for the left and right figure respectively. Hoeffding (``h'' in red); Pearson (``p'' in green); Spearman (``r'' in sky); Kendall (``k'' in blue); $L_n$ (``l'' in black).} 
\label{fig:indep.par} 
\end{figure}

\begin{figure}
[h!]
\centering 
\includegraphics[width=2.3in,height=2.5in]{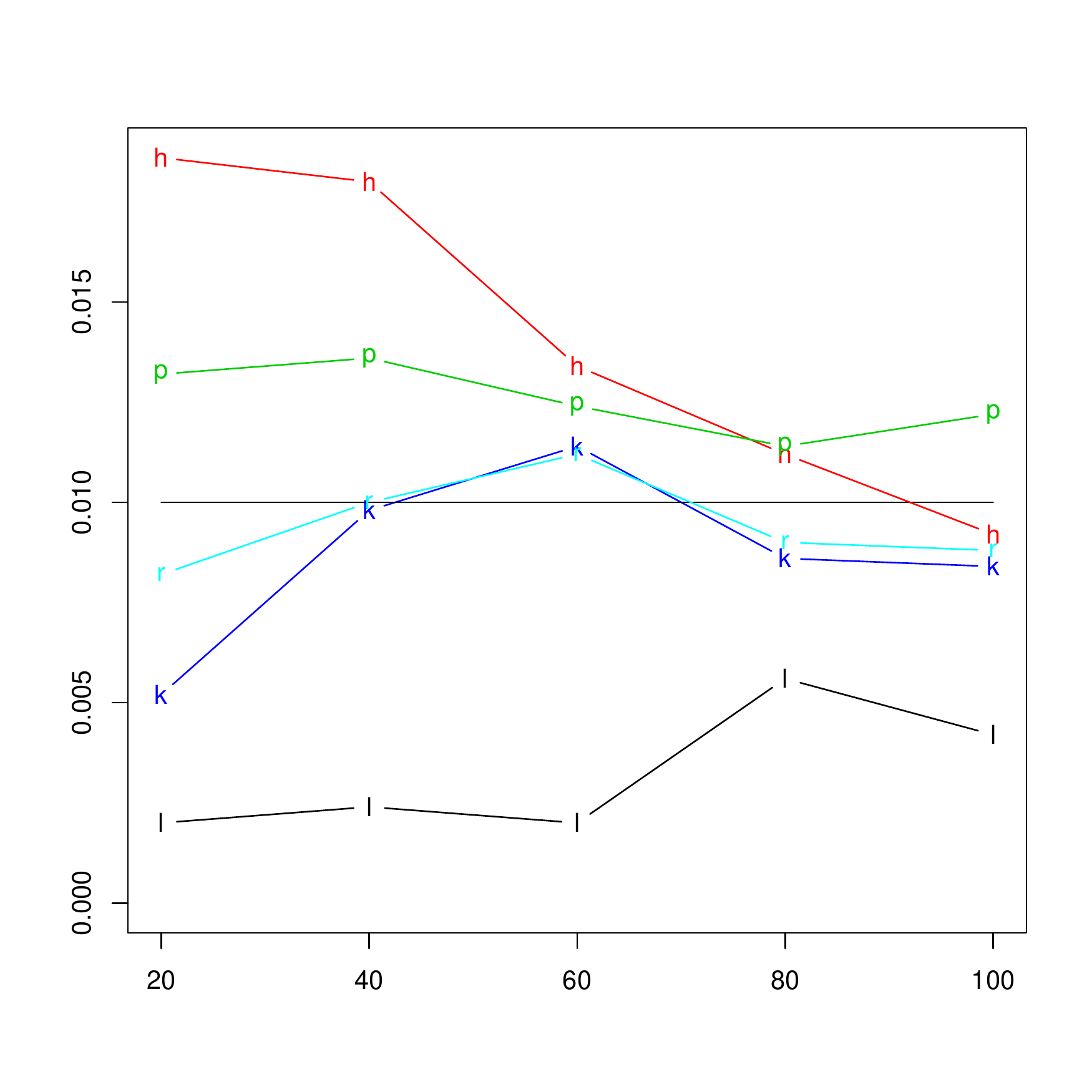}
\includegraphics[width=2.3in,height=2.5in]{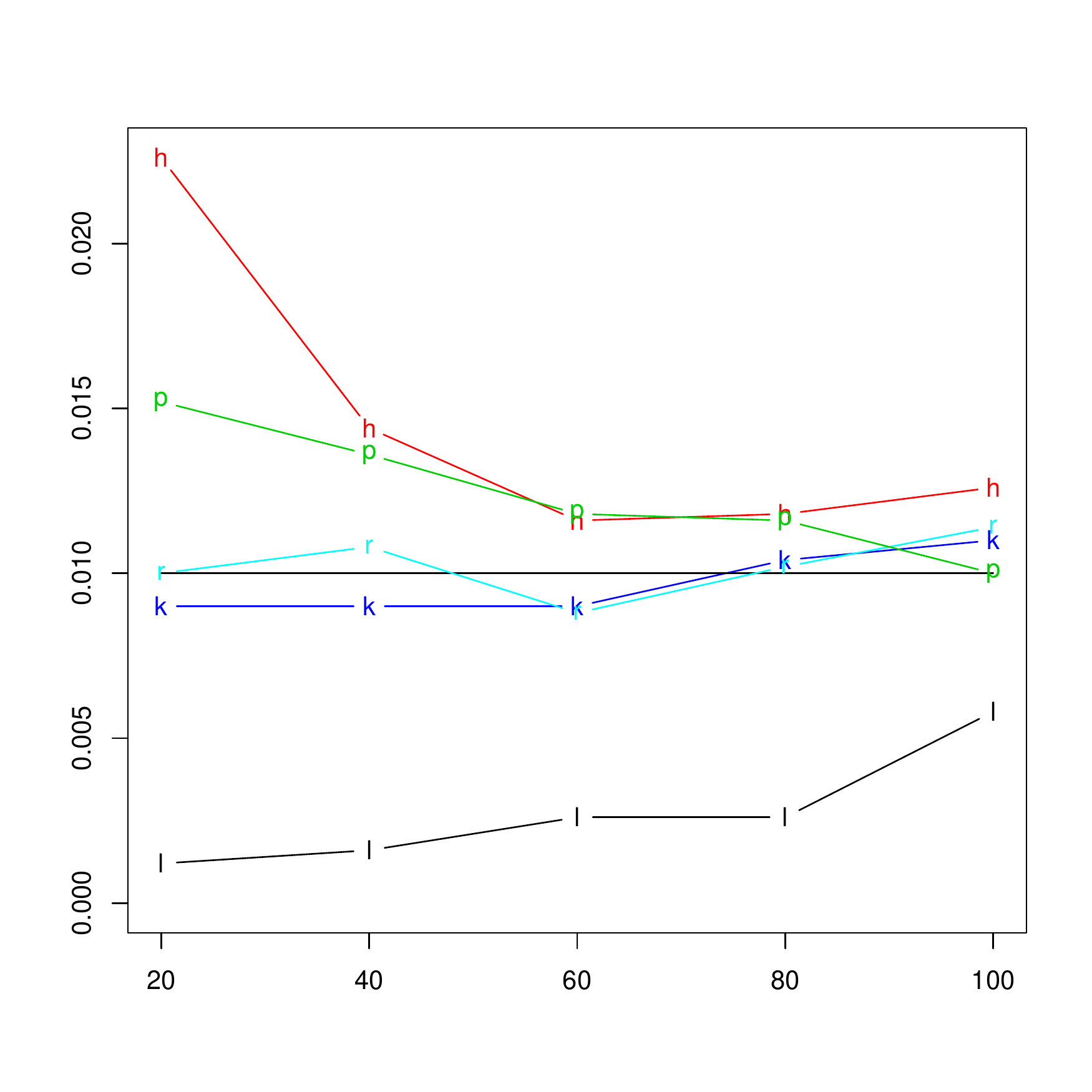}
 \caption{Sample size vs. empirical power Function at level 0.01 in the case of independent Weibull random variables with parameters $(1,0.25)$ and $(1,2)$ for the left and right figure respectively. Hoeffding (``h'' in red); Pearson (``p'' in green); Spearman (``r'' in sky); Kendall (``k'' in blue); $L_n$ (``l'' in black).} 
 \label{fig:indep.wei} 
\end{figure}

\begin{figure}
[h!]
\centering 
\includegraphics[width=2.3in,height=2.5in]{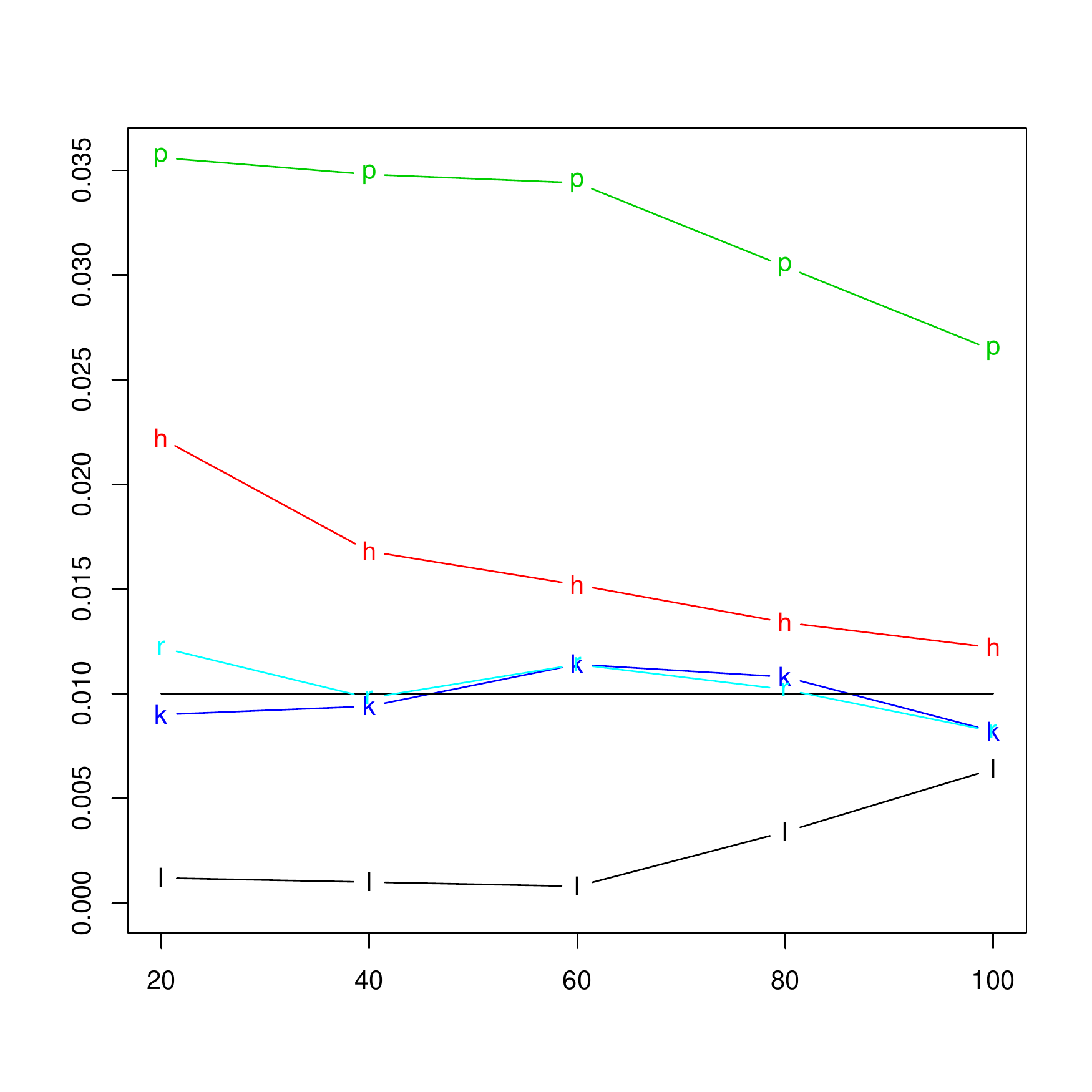}
\includegraphics[width=2.3in,height=2.5in]{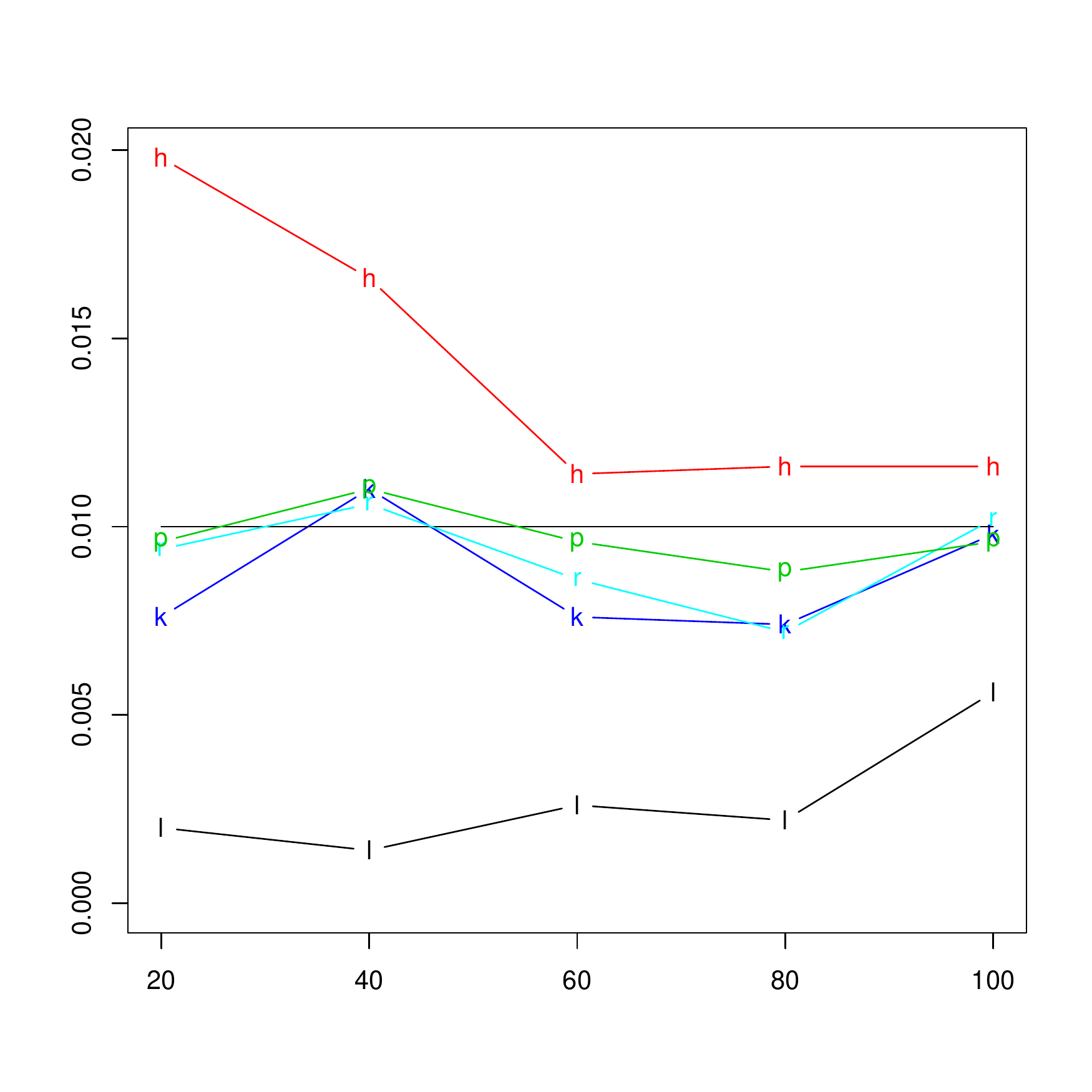}
 \caption{Sample size vs. empirical power Function at level 0.01 in the case of independent Student-t random variables with $1$ and $16$ degree of fredom for the left and right figure respectively. Hoeffding (``h'' in red); Pearson (``p'' in green); Spearman (``r'' in sky); Kendall (``k'' in blue); $L_n$ (``l'' in black).} 
 \label{fig:indep.t} 
\end{figure}

Figure \ref{fig:indep.par} shows the behavior of the empirical power functions, when $X$ and $Y$ have independent Pareto  marginals, with parameters of scale equal to 1; shape parameter equal to 0.25  for the picture on the left and shape parameter equal to 4,  for the picture on the right. For sample sizes going from 20 to 100, we can see that the only statistic with empirical power constantly lower than the level 0.01 is the $L_n$. We can see also that both, Pearson and Hoeffding tests can have empirical powers higher than 4 times the level 0.01. The other tests do not respect the significance levels for those sample sizes. A similar behavior can be seen in figure \ref{fig:indep.wei} under Weibull marginal distributions and under t-student marginal distributions in figure \ref{fig:indep.t}. This behavior can be seen in more details and for larger sample sizes on table \ref{tab:indep.div.01} for level $0.01$ and table \ref{tab:indep.div.05} for level $0.05$. 

\begin{table}
[h!]
\begin{center}
\begin{tabular}{llccccccc}
  \hline
Distribution & Test & 20 & 40 & 60 & 80 & 100 & 500 & 1000 \\ 
  \hline \hline
 & Spe & 0.011 & 0.011 & 0.010 & 0.009 & 0.010 & 0.010 & 0.008 \\ 
 & Ken & 0.010 & 0.011 & 0.010 & 0.009 & 0.009 & 0.010 & 0.008 \\ 
 Pareto(1,0.25) & Hoe & 0.023 & 0.017 & 0.013 & 0.013 & 0.012 & 0.009 & 0.010 \\ 
 & Pea & 0.051 & 0.032 & 0.025 & 0.019 & 0.017 & 0.003 & 0.003 \\ 
& Ln  & 0.001 & 0.002 & 0.001 & 0.004 & 0.005 & 0.007 & 0.007 \\ 
  \hline
 & Spe & 0.009 & 0.012 & 0.010 & 0.010 & 0.010 & 0.009 & 0.011 \\ 
 & Ken & 0.009 & 0.011 & 0.010 & 0.010 & 0.010 & 0.009 & 0.011 \\ 
  Pareto(1,4) & Hoe & 0.024 & 0.017 & 0.014 & 0.012 & 0.011 & 0.010 & 0.012 \\ 
 & Pea & 0.023 & 0.021 & 0.018 & 0.021 & 0.019 & 0.018 & 0.017 \\ 
 & Ln & 0.001 & 0.002 & 0.002 & 0.003 & 0.005 & 0.005 & 0.008 \\ 
\hline
\hline  
 & Spe & 0.009 & 0.011 & 0.010 & 0.009 & 0.010 & 0.011 & 0.010 \\ 
 & Ken & 0.008 & 0.010 & 0.011 & 0.009 & 0.009 & 0.011 & 0.010 \\ 
  Weibull(1,0.25) & Hoe & 0.021 & 0.016 & 0.013 & 0.012 & 0.010 & 0.011 & 0.008 \\ 
 & Pea & 0.015 & 0.012 & 0.012 & 0.011 & 0.012 & 0.011 & 0.011 \\ 
 & Ln & 0.001 & 0.002 & 0.002 & 0.004 & 0.005 & 0.008 & 0.004 \\ 
  \hline
 & Spe & 0.009 & 0.011 & 0.010 & 0.011 & 0.010 & 0.011 & 0.010 \\ 
 & Ken & 0.008 & 0.010 & 0.010 & 0.010 & 0.010 & 0.011 & 0.010 \\ 
  Weibull(1,2) & Hoe & 0.022 & 0.016 & 0.015 & 0.013 & 0.012 & 0.012 & 0.010 \\ 
& Pea & 0.013 & 0.012 & 0.012 & 0.012 & 0.012 & 0.011 & 0.010 \\ 
& Ln & 0.001 & 0.001 & 0.002 & 0.003 & 0.005 & 0.007 & 0.005 \\ 
  \hline
  \hline
 & Spe & 0.010 & 0.010 & 0.010 & 0.010 & 0.009 & 0.011 & 0.011 \\ 
 & Ken & 0.009 & 0.009 & 0.009 & 0.010 & 0.009 & 0.011 & 0.011 \\ 
  Student-t(1) & Hoe & 0.023 & 0.016 & 0.013 & 0.012 & 0.013 & 0.010 & 0.011 \\ 
 & Pea & 0.037 & 0.037 & 0.034 & 0.031 & 0.028 & 0.021 & 0.012 \\ 
 & Ln & 0.001 & 0.001 & 0.002 & 0.003 & 0.006 & 0.006 & 0.006 \\ 
  \hline
 & Spe & 0.009 & 0.010 & 0.009 & 0.009 & 0.011 & 0.011 & 0.009 \\ 
& Ken & 0.008 & 0.010 & 0.009 & 0.009 & 0.011 & 0.011 & 0.009 \\ 
  Student-t(16) & Hoe & 0.021 & 0.015 & 0.013 & 0.011 & 0.013 & 0.011 & 0.009 \\ 
 & Pea & 0.009 & 0.011 & 0.010 & 0.009 & 0.010 & 0.008 & 0.008 \\ 
 & Ln & 0.001 & 0.001 & 0.002 & 0.004 & 0.006 & 0.007 & 0.005 \\ 
  \hline
   \hline
\end{tabular}
\caption{Empirical power function at level 0.01, for pairs of i.i.d. random variables.}
\label{tab:indep.div.01}
\end{center}
\end{table}

\begin{table}[h!]
\begin{center}
\begin{tabular}{llccccccc}
  \hline
 Distribution & Test & 20 & 40 & 60 & 80 & 100 & 500 & 1000 \\ 
  \hline
  \hline
 & Spe & 0.051 & 0.051 & 0.051 & 0.051 & 0.051 & 0.053 & 0.046 \\ 
 & Ken & 0.047 & 0.049 & 0.050 & 0.051 & 0.050 & 0.053 & 0.046 \\ 
  Pareto(1,0.25) & Hoe & 0.079 & 0.065 & 0.060 & 0.056 & 0.057 & 0.054 & 0.046 \\ 
 & Pea & 0.058 & 0.037 & 0.028 & 0.022 & 0.019 & 0.004 & 0.003 \\ 
 & Ln & 0.019 & 0.024 & 0.014 & 0.022 & 0.028 & 0.046 & 0.038 \\ 
  \hline
 & Spe & 0.051 & 0.050 & 0.052 & 0.049 & 0.053 & 0.048 & 0.054 \\ 
& Ken & 0.049 & 0.049 & 0.052 & 0.049 & 0.052 & 0.048 & 0.054 \\ 
  Pareto(1,4) & Hoe & 0.083 & 0.065 & 0.060 & 0.054 & 0.058 & 0.048 & 0.055 \\ 
 & Pea & 0.056 & 0.051 & 0.050 & 0.049 & 0.047 & 0.048 & 0.052 \\ 
 & Ln & 0.019 & 0.023 & 0.015 & 0.021 & 0.031 & 0.048 & 0.038 \\ 
  \hline
  \hline
 & Spe & 0.047 & 0.051 & 0.049 & 0.051 & 0.050 & 0.051 & 0.049 \\ 
& Ken & 0.043 & 0.049 & 0.050 & 0.052 & 0.050 & 0.052 & 0.048 \\ 
Weibull(1,0.25) & Hoe & 0.077 & 0.065 & 0.058 & 0.056 & 0.054 & 0.050 & 0.047 \\ 
 & Pea & 0.047 & 0.049 & 0.049 & 0.047 & 0.046 & 0.049 & 0.053 \\ 
 & Ln & 0.019 & 0.022 & 0.015 & 0.021 & 0.028 & 0.048 & 0.034 \\ 
  \hline
 & Spe & 0.049 & 0.049 & 0.051 & 0.049 & 0.051 & 0.049 & 0.051 \\ 
 & Ken & 0.045 & 0.047 & 0.050 & 0.049 & 0.051 & 0.049 & 0.050 \\ 
  Weibull(1,2) & Hoe & 0.079 & 0.065 & 0.060 & 0.055 & 0.059 & 0.052 & 0.051 \\ 
& Pea & 0.051 & 0.048 & 0.049 & 0.048 & 0.050 & 0.054 & 0.043 \\ 
& Ln & 0.016 & 0.024 & 0.015 & 0.021 & 0.028 & 0.045 & 0.036 \\ 
  \hline
  \hline
 & Spe & 0.050 & 0.051 & 0.052 & 0.051 & 0.051 & 0.052 & 0.053 \\ 
 & Ken & 0.047 & 0.049 & 0.051 & 0.051 & 0.050 & 0.050 & 0.052 \\ 
  Student-t(1) & Hoe & 0.082 & 0.065 & 0.059 & 0.057 & 0.057 & 0.052 & 0.054 \\ 
 & Pea & 0.068 & 0.061 & 0.053 & 0.051 & 0.043 & 0.029 & 0.018 \\ 
 & Ln & 0.019 & 0.024 & 0.014 & 0.020 & 0.028 & 0.050 & 0.034 \\ 
  \hline
 & Spe & 0.048 & 0.049 & 0.050 & 0.048 & 0.052 & 0.049 & 0.046 \\ 
 & Ken & 0.045 & 0.047 & 0.051 & 0.049 & 0.052 & 0.048 & 0.046 \\ 
  Student-t(16) & Hoe & 0.081 & 0.063 & 0.060 & 0.056 & 0.057 & 0.048 & 0.045 \\ 
 & Pea & 0.050 & 0.050 & 0.050 & 0.049 & 0.050 & 0.051 & 0.045 \\ 
 & Ln & 0.020 & 0.025 & 0.014 & 0.022 & 0.031 & 0.050 & 0.033 \\ 
  \hline
   \hline
\end{tabular}
\caption{Empirical power function at level 0.05, for pairs of i.i.d. random variables. }
\label{tab:indep.div.05}
\end{center}
\end{table}

Figure  \ref{fig:indep.wei} shows the empirical power functions assuming a Weibull for each variable with scale parameter equal to 1 and shape parameter equal to 0.25 on the left and on the right, the scale parameter is equal to 1 and the shape parameter is equal to 2. Figure \ref{fig:indep.t} shows the empirical power functions under Student-t marginals with one degree of freedom on the left and 16 degrees of freedom on the right.

\subsection{Dependence}

\begin{figure}
[h!] 
\centering 
\includegraphics[width=4in,height=3in]{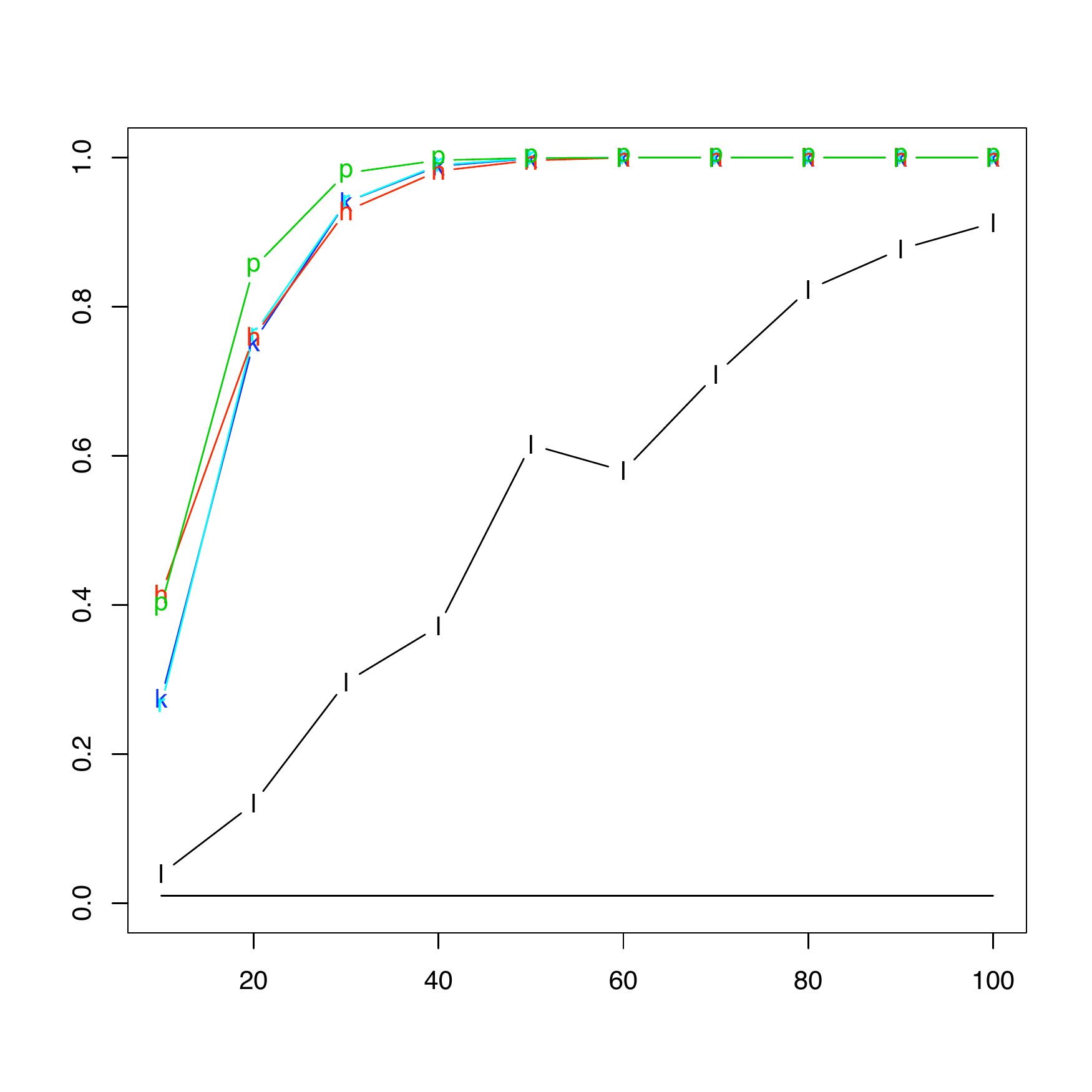} \caption{Sample size vs. empirical power function at level 0.01. Bivariate case, with $N(0,1)$ marginals and $\rho = 0.7.$ Hoeffding (``h'' in red); Pearson (``p'' in green); Spearman (``r'' in sky); Kendall (``k'' in blue); $L_n$ (``l'' in black). }
 \label{fig:depnormal} 
\end{figure}

Figure \ref{fig:depnormal}
shows the power functions at level $0.01$ in the bivariate case, with standard normal marginal distributions and correlation coefficient equal to 0.7. As is expected for the normal case, the Hoeffding and Pearson tests have the highest power functions. Table \ref{tab:depnormal} shows the power for level $0.05$.

\begin{table}[h!]
\begin{center}
\begin{tabular}{rrrrrr}
  \hline
 & Spearman & Kendall & Hoeffding & Pearson & $L_n$  \\
  \hline
10 & 0.566 & 0.555 & 0.590 & 0.681 & 0.215  \\
  20 & 0.914 & 0.910 & 0.893 & 0.956 & 0.351  \\
  30 & 0.988 & 0.988 & 0.979 & 0.996 & 0.559  \\
  40 & 0.999 & 0.999 & 0.998 & 1.000 & 0.618  \\
  50 & 1.000 & 1.000 & 1.000 & 1.000 & 0.824  \\
  60 & 1.000 & 1.000 & 1.000 & 1.000 & 0.779  \\
  70 & 1.000 & 1.000 & 1.000 & 1.000 & 0.877  \\
  80 & 1.000 & 1.000 & 1.000 & 1.000 & 0.932  \\
  90 & 1.000 & 1.000 & 1.000 & 1.000 & 0.960  \\
  100 & 1.000 & 1.000 & 1.000 & 1.000 & 0.973  \\
   \hline
\end{tabular}
\caption{Empirical power function at level 0.05. Bivariate case with $N(0,1)$  marginal distributions and $\rho = 0.7$. }
\label{tab:depnormal}
\end{center}
\end{table}

\begin{figure}
[h!] 
\centering 
\includegraphics[width=4in,height=3in]{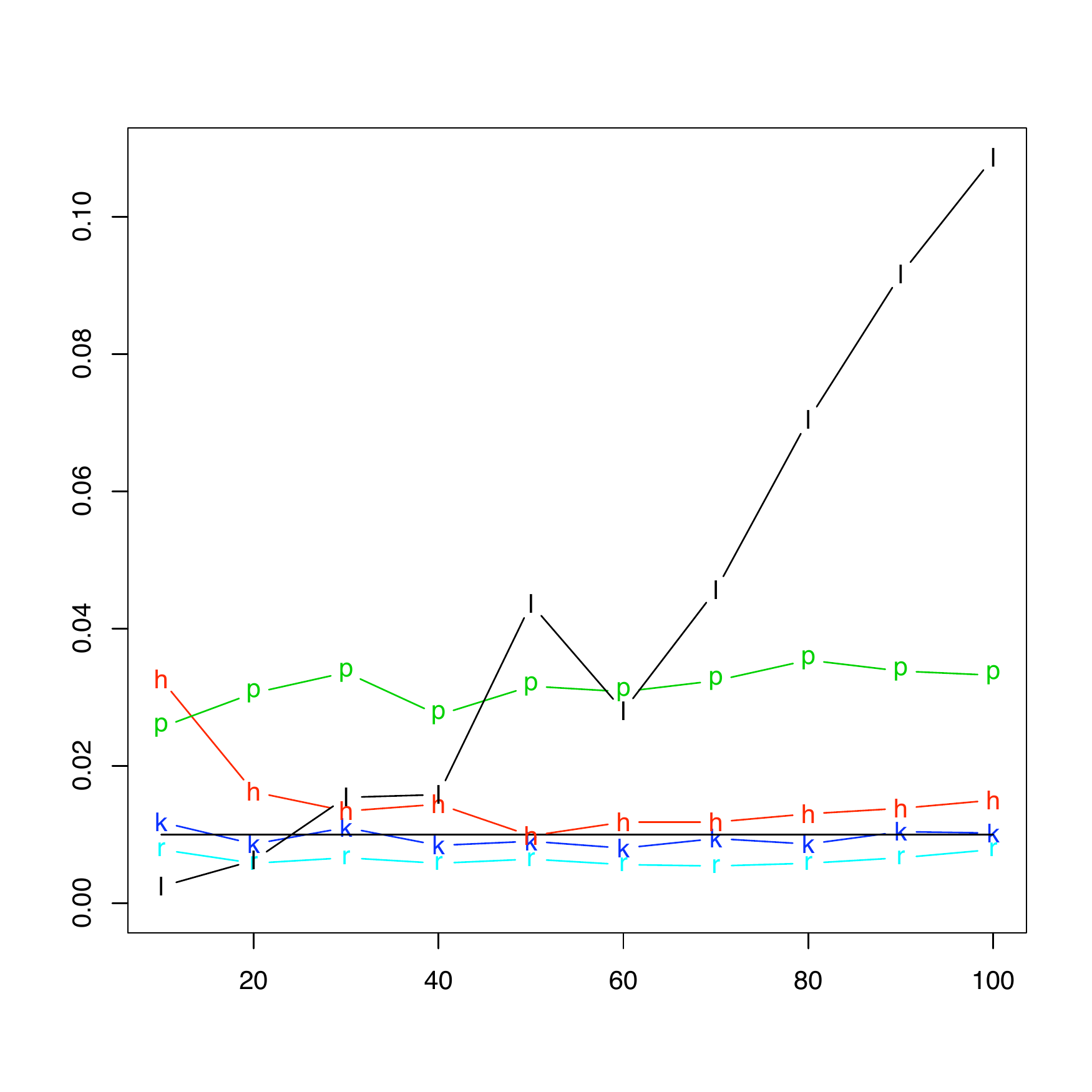} \caption{Sample size vs. empirical power function at level 0.01. Mixture (50-50) of two bivariate distributions with standard normal marginals, one with $\rho = 0.7$ and the other with $\rho = -0.7$. Hoeffding (``h'' in red); Pearson (``p'' in green); Spearman (``r'' in sky); Kendall (``k'' in blue); $L_n$ (``l'' in black). } 
\label{fig:depcrossnormal} 
\end{figure}

Figure \ref{fig:depcrossnormal}
shows the power empirical functions at level $0.01$ in the case of a mixture (50-50) of two bivariate with standard normal marginal distributions one with positive correlation $0.7$ and the other with negative correlation $-0.7$. In this case, the $L_n$ test has the highest power function.  Table \ref{tab:dep.mix.05} shows the power for level $0.05$.

\begin{table}[h!]
\begin{center}
\begin{tabular}{llccccccc}
  \hline
 $\rho$ & Test & 20 & 40 & 60 & 80 & 100 & 500 & 1000 \\ 
  \hline
   & Spe & 0.007 & 0.008 & 0.008 & 0.008 & 0.009 & 0.007 & 0.006 \\ 
     & Ken & 0.008 & 0.009 & 0.010 & 0.010 & 0.010 & 0.009 & 0.008 \\ 
    0.5 & Hoe & 0.018 & 0.014 & 0.012 & 0.011 & 0.012 & 0.014 & 0.022 \\ 
     & Pea & 0.016 & 0.018 & 0.021 & 0.020 & 0.022 & 0.020 & 0.021 \\ 
     & Ln & 0.002 & 0.004 & 0.005 & 0.013 & 0.025 & 0.121 & 0.213 \\ 
  \hline
    & Spe & 0.009 & 0.007 & 0.007 & 0.007 & 0.007 & 0.006 & 0.006 \\ 
     & Ken & 0.010 & 0.010 & 0.010 & 0.010 & 0.010 & 0.009 & 0.008 \\ 
    0.6 & Hoe & 0.018 & 0.013 & 0.012 & 0.011 & 0.013 & 0.025 & 0.104 \\ 
     & Pea & 0.024 & 0.026 & 0.025 & 0.027 & 0.027 & 0.028 & 0.025 \\ 
     & Ln & 0.003 & 0.006 & 0.011 & 0.030 & 0.048 & 0.301 & 0.528 \\ 
    \hline
  & Spe & 0.006 & 0.006 & 0.006 & 0.006 & 0.007 & 0.006 & 0.007 \\ 
   & Ken & 0.009 & 0.009 & 0.009 & 0.009 & 0.010 & 0.009 & 0.010 \\ 
    0.7 & Hoe & 0.017 & 0.014 & 0.012 & 0.012 & 0.013 & 0.109 & 0.865 \\ 
   & Pea & 0.028 & 0.030 & 0.033 & 0.033 & 0.033 & 0.033 & 0.036 \\ 
   & Ln & 0.004 & 0.013 & 0.024 & 0.061 & 0.104 & 0.655 & 0.902 \\   
  \hline
  & Spe & 0.006 & 0.006 & 0.005 & 0.005 & 0.005 & 0.007 & 0.005 \\ 
   & Ken & 0.008 & 0.009 & 0.008 & 0.009 & 0.009 & 0.010 & 0.008 \\ 
    0.8 & Hoe & 0.016 & 0.012 & 0.013 & 0.015 & 0.018 & 0.889 & 1.000 \\ 
   & Pea & 0.037 & 0.040 & 0.042 & 0.041 & 0.044 & 0.046 & 0.046 \\ 
   & Ln & 0.007 & 0.029 & 0.064 & 0.165 & 0.251 & 0.951 & 0.999 \\ 
  \hline
   & Spe & 0.005 & 0.004 & 0.005 & 0.005 & 0.005 & 0.005 & 0.004 \\ 
   & Ken & 0.007 & 0.007 & 0.007 & 0.007 & 0.007 & 0.009 & 0.007 \\ 
    0.9 & Hoe & 0.012 & 0.014 & 0.021 & 0.035 & 0.058 & 1.000 & 1.000 \\ 
   & Pea & 0.047 & 0.052 & 0.054 & 0.054 & 0.054 & 0.053 & 0.054 \\ 
   & Ln & 0.017 & 0.103 & 0.244 & 0.501 & 0.673 & 1.000 & 1.000 \\ 
  \hline
\end{tabular}
\caption{Empirical power function at level 0.01.   Mixture (50-50) of bivariate distributions with $N(0,1)$ marginals, one with $\rho$ and the other with $-\rho.$}
\label{tab:dep.mix.01}
\end{center}
\end{table}

\begin{table}[h!]
\begin{center}
\begin{tabular}{llccccccc}
  \hline
 $\rho$ & Test & 20 & 40 & 60 & 80 & 100 & 500 & 1000 \\ 
  \hline
  & Spe & 0.042 & 0.043 & 0.044 & 0.042 & 0.042 & 0.041 & 0.040 \\ 
   & Ken & 0.044 & 0.049 & 0.050 & 0.049 & 0.050 & 0.047 & 0.047 \\ 
    0.5 & Hoe & 0.070 & 0.059 & 0.054 & 0.054 & 0.051 & 0.074 & 0.143 \\ 
   & Pea & 0.071 & 0.074 & 0.078 & 0.075 & 0.081 & 0.078 & 0.078 \\ 
   & Ln & 0.021 & 0.026 & 0.026 & 0.047 & 0.073 & 0.346 & 0.456 \\ 
  \hline
   & Spe & 0.043 & 0.041 & 0.040 & 0.040 & 0.040 & 0.037 & 0.037 \\ 
   & Ken & 0.048 & 0.049 & 0.049 & 0.050 & 0.049 & 0.047 & 0.049 \\ 
   0.6 & Hoe & 0.068 & 0.057 & 0.053 & 0.054 & 0.054 & 0.160 & 0.579 \\ 
   & Pea & 0.086 & 0.090 & 0.088 & 0.089 & 0.094 & 0.093 & 0.092 \\ 
   & Ln & 0.024 & 0.033 & 0.040 & 0.088 & 0.120 & 0.605 & 0.782 \\ 
  \hline
   & Spe & 0.038 & 0.036 & 0.035 & 0.036 & 0.038 & 0.035 & 0.034 \\ 
   & Ken & 0.044 & 0.045 & 0.048 & 0.049 & 0.049 & 0.048 & 0.046 \\ 
    0.7 & Hoe & 0.063 & 0.054 & 0.056 & 0.061 & 0.068 & 0.620 & 0.999 \\ 
   & Pea & 0.099 & 0.102 & 0.105 & 0.109 & 0.110 & 0.107 & 0.108 \\ 
   & Ln & 0.031 & 0.050 & 0.076 & 0.154 & 0.224 & 0.881 & 0.977 \\ 
  \hline
   & Spe & 0.036 & 0.032 & 0.033 & 0.032 & 0.032 & 0.035 & 0.034 \\ 
   & Ken & 0.045 & 0.045 & 0.045 & 0.047 & 0.045 & 0.048 & 0.050 \\ 
    0.8 & Hoe & 0.060 & 0.057 & 0.067 & 0.083 & 0.101 & 1.000 & 1.000 \\ 
   & Pea & 0.117 & 0.125 & 0.120 & 0.121 & 0.123 & 0.129 & 0.128 \\ 
   & Ln & 0.046 & 0.099 & 0.165 & 0.321 & 0.435 & 0.993 & 1.000 \\ 
  \hline
   & Spe & 0.031 & 0.030 & 0.030 & 0.028 & 0.030 & 0.031 & 0.029 \\ 
   & Ken & 0.039 & 0.042 & 0.043 & 0.043 & 0.043 & 0.044 & 0.040 \\ 
    0.9 & Hoe & 0.060 & 0.079 & 0.124 & 0.212 & 0.357 & 1.000 & 1.000 \\ 
   & Pea & 0.135 & 0.139 & 0.139 & 0.141 & 0.142 & 0.140 & 0.137 \\ 
   & Ln & 0.093 & 0.257 & 0.443 & 0.702 & 0.826 & 1.000 & 1.000 \\ 
   \hline
\end{tabular}
\caption{Empirical power function at level 0.05.  Mixture (50-50) of bivariate distributions with $N(0,1)$ marginals, one with $\rho$ and the other with $-\rho.$}
\label{tab:dep.mix.05}
\end{center}
\end{table}

Tables \ref{tab:dep.mix.01} and \ref{tab:dep.mix.05} show the tendencies of the power function  under the mixture (50-50) of bivariate distributions with standard normal marginal distributions, one with a correlation coefficient equal to $\rho$ and the other with a correlation coefficient equal to $-\rho ;$ $\rho$ taking the values 0.5,0.6,0.7 and 0.9. In both tables, the $L_n$ test achieves the higher values in the power function, when the sample size grows. 

\begin{figure}
[h!]
\centering 
\includegraphics[width=2.3in,height=2in]{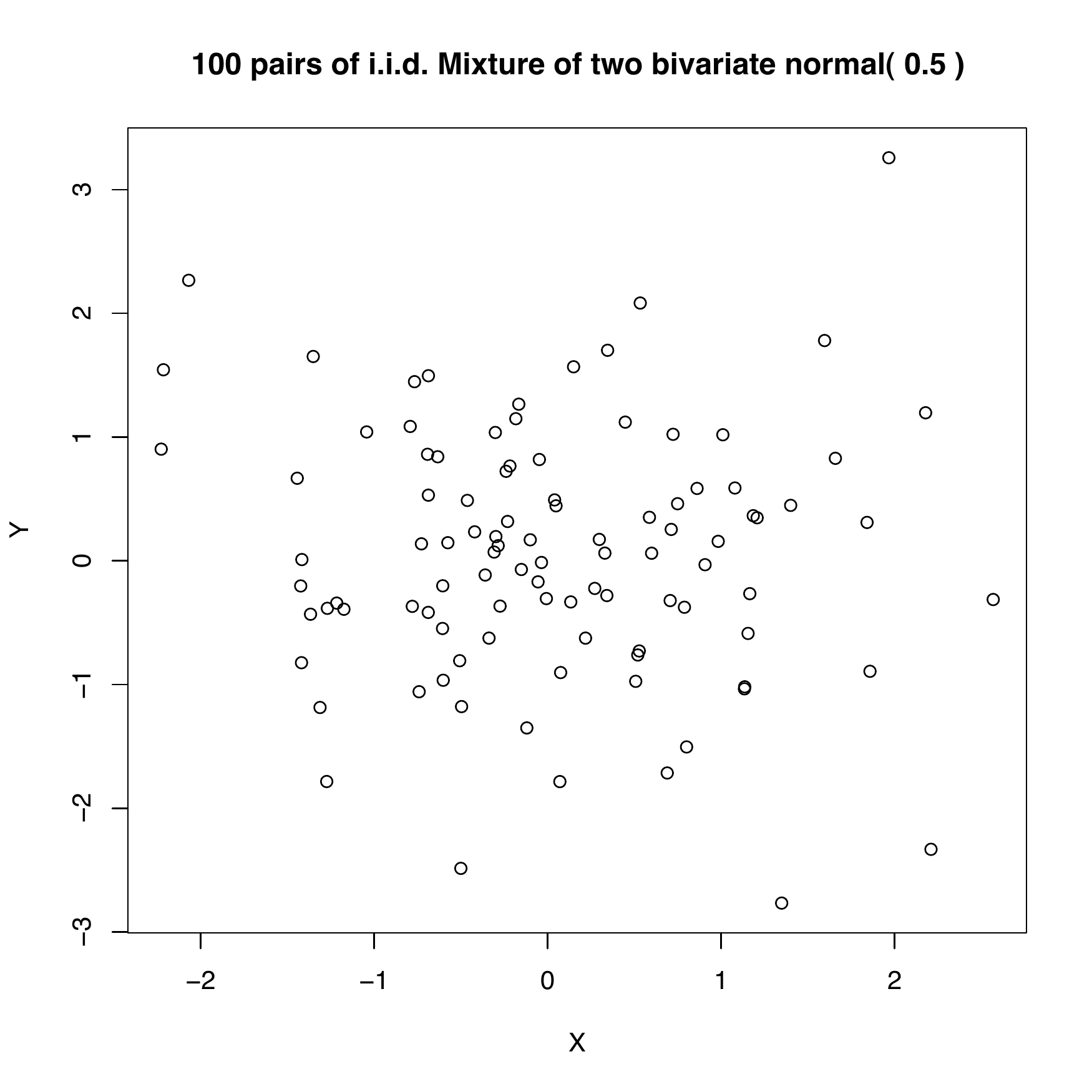}
\includegraphics[width=2.3in,height=2in]{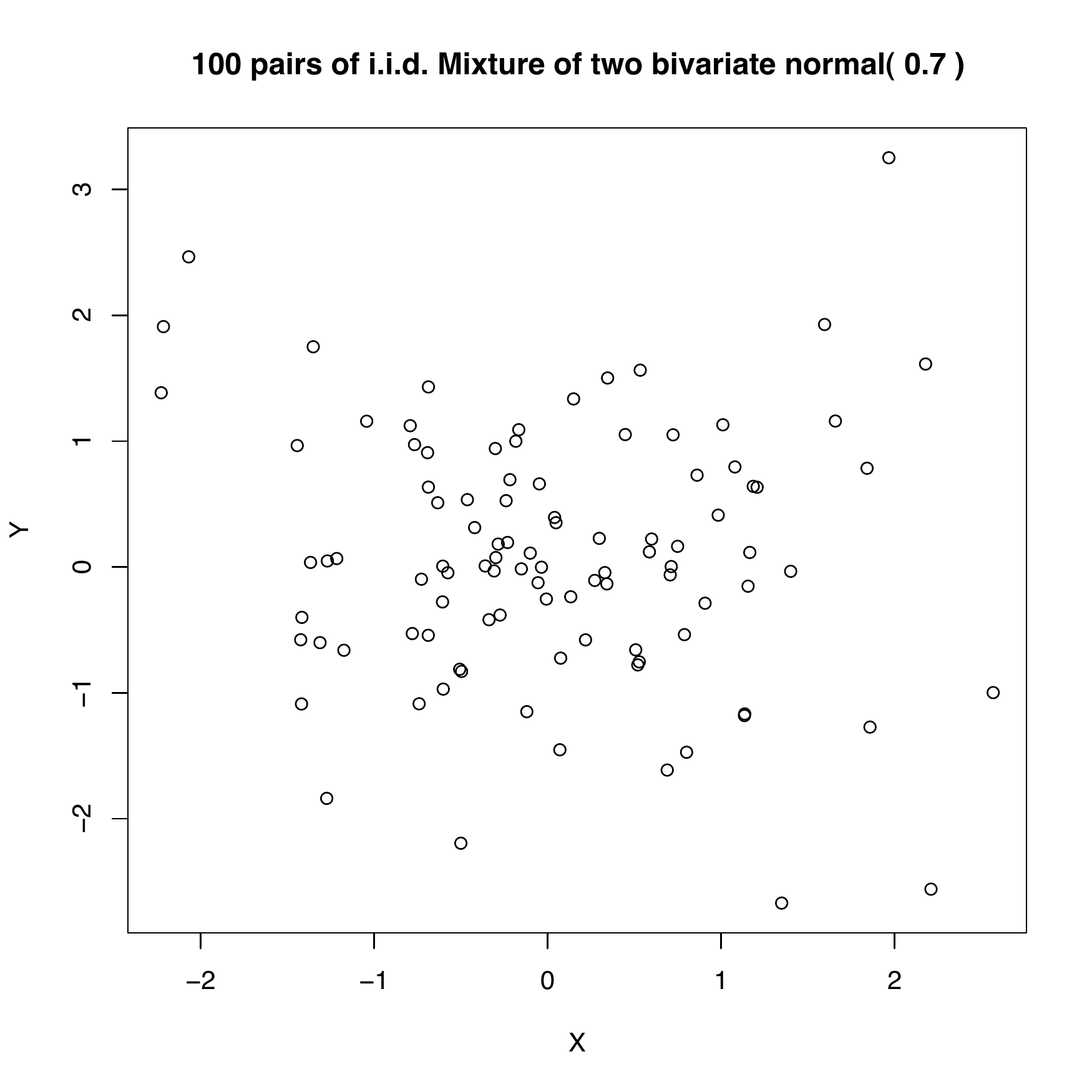}
 \caption{Plot of the sample. Mixture (50-50) of bivariate distributions with $N(0,1)$ marginals, one with $\rho$ and the other with $-\rho.$ On the left $\rho=0.5,$ on the right $\rho=0.7.$} 
 \label{mixture5050} 
\end{figure}

\section{Conclusions}\label{conclu} Under the assumption of independence, by construction, the $L_n$ test respects the significance level, as showed in the simulation study. In contrast, for moderate sample size (between 20 and 100) we report the lack of control of the power function for Pearson and Hoeffding test in the case of heavy tailed marginal distributions, like Weibull, Pareto and t-student (with a small degree of freedom). This means that, for small $n$, when we do not have information about the normality of the marginal distributions, it is recommended the procedure $L_n.$  We enphasize that even when the sample size is equal to 1000 our simulations show  the lack of control of the significance level for Spearman, Kendall, Hoeffding and Pearson tests under the assumption of heavy tailed distributions. \\
Under the assumption of normality with high correlation coefficient, the power function of $L_n$ test grows with the sample size, but the recommended procedure is Pearson, as was expected. $L_n$ could be compared with Pearson from a sample size equal to 80.  Our procedure reports the remarkable behavior of its power function when applied in mixtures of bivariate normal distributions. We observed that Pearson can not detect the dependence even with a high value of correlation and $L_n$ is recommended in that case. Considering the mixture given by 50\% of bivariate distribution of standard normal marginals with correlation coefficient equal to $\rho$ and 50\% of bivariate distribution of standard normal marginals with correlation coefficient equal to $-\rho,$ we report that $L_n$ shows a growing power function that is much higher when $\rho$ grows. See for illustration the plots $X$ vs $Y$ in two cases, $\rho=0.5$ and $\rho=0.7$, figure \ref{mixture5050}. In those cases the other tests appear less powerful than $L_n.$    

\appendix

\section{Background. Statistical tests}\label{backg} The Pearson test, Kendall test, Spearman test and Hoeffding test are tests for association between paired samples. 
Pearson test checks if $\rho=0,$ where $\rho$ is the correlation between the two variables $X$ and $Y.$ The test is supported by the Student-$t$ statistic and based on Pearson's product moment correlation coefficient $r,$ which is the correlation between the two variables in the sample. $t=r\sqrt{\frac{n-2}{1-r^2}}$ has the $t$ distribution with $n-2$ degrees of freedom when the samples follow the independent normal distribution. \\
The SpearmanÕs rank correlation coefficient uses the percentiles of a distribution to define the statistic, the formal expression of this coefficient is given by $\rho_s=12\int \int [H(x,y)-F(x)G(y)]dF(x)dG(y).$ This measure is cheked by the SpearmanÕs rank test. $(X,Y)$ is said to be positively quadrant dependent, if $H(x,y)-F(x)G(y) \geq 0\,\, \forall x,y.$ So, $\rho_s$ represents an average which measures the positive quadrant dependence. Where the average is taken with respect to the marginal distributions of $X$ and $Y.$ The sample version of $\rho_s$ is given by $r_{s}=\frac{12}{n(n^2-1)}\sum_{i=1}^n(rank(x_i)-\frac{n+1}{2})(rank(y_i)-\frac{n+1}{2}).$ \\
The Kendall Tau is a measure of the condition ``total positivity of order two''. A pair of random variables $(X,Y)$ with an absolutely continuous distribution function $H$ is said to be totally positive of order two if the joint density function $h(x,y)$ satisfies $h(x_2,y_2)h(x_1,y_1)-h(x_1,y_2)h(x_2,y_1) \geq 0, $ whenever $x_1 <x_2$ and $y_1 < y_2.$ So, the Kendall Tau coefficient defined by $\tau=2\int_{-\infty}^{\infty}\int_{-\infty}^{\infty}\int_{-\infty}^{y_2}\int_{-\infty}^{x_2}[h(x_2,y_2)h(x_1,y_1)-h(x_1,y_2)h(x_2,y_1)]dx_1dy_1dx_2dy_2$ measures this property. The sample version of $\tau,\,\,\tau_s$ is defined as the product moment correlation of ``signs of concordance'', and this is the statistic used by the Kendall Tau test. Formally, we define the $s$ function as $s(x)=1$ when $x>0,$ $s(x)=\frac{1}{2}$ when $x=0$ and $s(x)=-1$ when $x<0.$ We have, $\tau_s=\frac{1}{n(n-1)}\sum \sum_{i \neq j} s(X_i-X_j)s(Y_i-Y_j),$ $\tau_s$ explains how well the two sequences follow a monotone order.\\
To compute the $P$-values for each method, we use the ``cor.test'' function, available in the ``stat'' package from R-project. More details about each test may be found in \citet{Hollander1973}.\\
The last method is based on the HoeffdingÕs measure, proposed by \citet{Hoeff1948} and supported by the notion of distance between two distributions. Formally, the measure is given by $\Delta=\int \int [H(x,y)-F(x)G(y)]^2dH(x,y).$ This measure is appropriate only when $H$ is absolutely continuous. The sample measure $\Delta_n$ is defined by 
$\Delta_n=A-2(n-2)B+(n-2)(n-3)C\frac{(n-5)!}{n!} $
where 
\begin{eqnarray*}
A&=&\sum_{i=1}^2(rank(x_i)-1)(rank(x_i)-2)(rank(y_i)-1)(rank(y_i)-2),\\
B&=&\sum_{i=1}^n(rank(x_i)-2)(rank(y_i)-2)T_i,\\
C&=&\sum_{i=1}^nT_i(T_i-1),\,\,T_i=\left\{ j : x_j < x_i\,\mbox{and}\,y_j<y_i\right\}. 
\end{eqnarray*}
In this last case, to compute the $P$-values, we use the ``hoeffd'' function, available in the ``Hmisc'' package from R-project.


\end{document}